\documentclass[a4paper]{jpconf}
\usepackage{graphicx}
\begin{document}
\title{Connecting the numerical and analytical ionization times for quantum dots in semiconductor wires driven by alternating field}

\author{D.V. Khomitsky}

\address{Department of Physics, National Research Lobachevsky State University of Nizhny Novgorod, 603950 Gagarin Avenue 23, Nizhny Novgorod, Russian
Federation}

\ead{khomitsky@phys.unn.ru}

\begin{abstract}
Recent numerical results for ionization of quantum dots by periodic electric field during the electric dipole spin resonance are compared with known analytical approaches. It is found that in finite-length quantum wires the numerical ionization rate is slower than the analytical one, especially for low driving field when the confinement potential strongly affects the dynamics. Still, the analytically predicted ionization times are in satisfactory correspondence with mean energy threshold times when the energy crosses the border between localized and continuum states.
\end{abstract}

\section{Introduction}

The problem of quantitative analytical description of the ionization of atoms by constant and alternating electric field has been a challenge for decades \cite{DK1985,DK2000}. Since the pioneering work by Keldysh \cite{Keldysh1964} it has been found that the ionization rate is very sensitive to the electric field strength $F=|e|{\cal E}$ in all regimes of ionization including the periodic driving. For slowly varying fields when the adiabaticity parameter $\gamma=\hbar \omega \kappa/F$ is lower than $1$ where $\omega$ is the driving frequency and $\kappa$ is the typical inverse localization length for confined state, the dependence is of exponential nature with the form $\sim \exp(-2 F_0/3 F)$, $F_0$ being a typical electric field of the confining potential. Such exponential dependence has been known for decades for the models of the tunneling rates for the barriers affected by static field. However, the calculations of an accurate pre-exponential factor needed for comparison of analytics with experimental or numerical results face certain difficulties and are essentially model-dependent. A great variety of models has been developed within various approximations, especially for the problems from the atomic physics \cite{DK1985,DK2000}. Usually the atomic potentials considered there are three-dimensional, but one-dimensional models with analytic solutions for the ionization rates produced by alternating field have also been proposed \cite{Perelomov1966,Demi1996}. They are of special interest for the physics of low-dimensional structures and, in particular, for the problem of ionization of quantum wells formed by one-dimensional potential \cite{Demi1996}. In these structures with many examples such as nanowires with gate-defined quantum dots the ionization rate still obeys some universal rules for simple models like delta-function potential \cite{Perelomov1966} but its pre-exponential factor is again model and geometry-dependent \cite{Demi1996}. If one compares the numerical results for atomic ionization with analytical approximations \cite{Bauer1999} then it becomes clear that none of the existing analytical approaches describes the numerical results with good accuracy in all range of parameters. This usually means that the correspondence better than the same order of magnitude is hardly achievable. Instead, an empirical model for atomic ionization rates have been proposed fitting the numerical data \cite{Bauer1999}.  

In our recent papers \cite{KLS2019,KLS2020} we have considered the problem of electron driven dynamics in one-dimensional quantum dots formed either by shallow donor potential \cite{KLS2019} or by deeper electrostatic gate-defined well \cite{KLS2020} in InSb-based nanowires with large spin-orbit coupling (SOC). We were interested in the development of electric dipole spin resonance (EDSR) combined with the tunneling to continuum caused by alternating electric field. It was found numerically in \cite{KLS2020} that the tunneling is very sensitive to the electric field strength which resembles the above mentioned results on atomic ionization. However, the quantitative correspondence with the analytic approximations is always a factor that needs clarification.

Here we analyze the numerical localization probability decay and associated ionization times for our multilevel quantum dot model from \cite{KLS2020} and connect it with most relevant analytical approximation. It is found that although the precise correspondence between numerical and analytical results in terms of ionization times is still hardly achievable, satisfactory quantitative conclusions can be made which highlight the connections between localization probability and mean energy evolution.

\section{Hamiltonian and parameters of quantum well for numerical evolution}

In our model developed in \cite{KLS2019,KLS2020} we considered a nanowire oriented along the $x-$ axis, where the effective mass Hamiltonian can be described as 

\begin{equation}
H_0=\frac{\hbar^{2}k^{2}}{2m}+U(x)+
\alpha \sigma_{y}k+\frac{\Delta}{2}\sigma_{z}. 
\label{h0}
\end{equation}

Here $m=0.0136 m_0$ is the electron effective mass for InSb nanowire, and $k=-i\partial/\partial x$ is the wave vector operator. 
We took a simple model of a gate-defined confinement potential $U(x)=-U_0/(\cosh^2(x/d))$. In the present paper we shall focus on the deep multilevel well considered in \cite{KLS2020} where we used the parameters $d=50$ nm and $U_0=27$ meV which corresponded to the formation of five discrete levels with energies $E_{n}^{(0)}<0$ by two first terms in (\ref{h0}). The distance between the lowest level $E_{1}^{(0)}$ and the next level $E_{2}^{(0)}$ provides a natural frequency scale $\omega_0=1.343 \cdot 10^{13}$ 1/s and the associated period gives the time scale $T_0=0.486$ ps. The third and fourth terms in (\ref{h0}) are responsible for Rashba spin-orbit coupling and Zeeman splitting in the static magnetic field. In this paper we took examples from \cite{KLS2020} for $\alpha=5$ $\rm{meV \cdot nm}$ and the static magnetic field $(0,0,B_z)$ with $Bz=0.447$ T. For the electron g-factor $g=-50.6$ in InSb nanowire this provides the Zeeman splitting $\Delta$ in the last term of (\ref{h0}) giving  the splitting $E_2-E_1=1.313$ meV corresponding to the driving frequency $\omega=2 \cdot 10^{12}$ 1/s. We label the spin-resolved levels and eigenfunctions as $E_n$ and $\phi_n(x)$, respectively. The ground state has the energy $E_1=-22.7$ meV. The value of $|E_1|$ is usually labeled as "ionization potential" in the ionization theory \cite{DK1985,DK2000}. The presence of continuum states has been taken into account in \cite{KLS2019,KLS2020} by finding the numerically accurate eigenstates of (\ref{h0}) with positive energies. They formed a dense discrete set of states since the nanowire has been considered to be of finite length $2L=16$ mkm in \cite{KLS2020} which is longer than in most of experiments. Below we shall see that the finite nanowire length leads to certain corrections for ionization rates compared to ideal infinite continuum. 

The dynamics is considered for the electron initially located on the ground level $E_1$ and driven by the uniform electric field $(F(t),0,0)$  where $F(t)=|e| {\cal E}(t)$, producing the scalar potential

\begin{equation}
V(x,t)=Fx\sin \omega t.
\label{vxt}
\end{equation}

In \cite{KLS2019,KLS2020} we solved the nonstationary Schr\"{o}dinger equation with Hamiltonian $H=H_0+V(x,t)$ obtained from (\ref{h0}) and (\ref{vxt}), and represented the wavefunction as a sum over all eigenstates of $H_0$ with time-dependent coefficients: $\Psi(x,t)=\sum_n C_n(t) \phi_n(x) \exp(-i E_n t/\hbar)$. The unitary evolution of $C_n(t)$ with the initial condition $C_1(0)=1$ has been obtained in \cite{KLS2020} by numerical Cayley procedure via the standard techniques.

Our focus in \cite{KLS2020} was on the evolution of observables like the electron mean spin projections, mean energy, coordinate, etc. The field-induced tunneling to continuum has been described by the localization probability 

\begin{equation}
P_{\rm{loc}}(t)=\sum_{n\rm{(loc)}} |C_n(t)|^2,
\label{probloc}
\end{equation}

where the sum is taken over the localized basis states with the energy $E_n<0$. The time dependence of $P_{\rm{loc}}(t)$ can be considered as the evolution of the electron probability to stay bound to the confining potential, and the characteristic rate and time of its decay can be viewed as the ionization rate and time, respectively. Below we shall analyze the numerical results for $P_{\rm{loc}}(t)$ obtained in \cite{KLS2020} for various driving field strength on the time interval $(0, \ldots ,650) T_0$ which ends below $0.4$ ns allowing us to ignore the disorder or temperature-induced relaxation and decoherence processes on such short time scale.

\section{Numerical and analytical results for ionization times}

The ionization rate $w(t)$ can be obtained from the time dependence of numerically found localization probability (\ref{probloc}) by considering the exponential fit of the form \cite{Bauer1999}:

\begin{equation}
P_{\rm{loc}}(t)=P_0 \exp \left(- \Gamma (t) \right), \quad
\Gamma(t)=\int_{t_0}^t w(t') dt'.
\label{gamma}
\end{equation}

\begin{figure}[h]
\includegraphics[width=115mm]{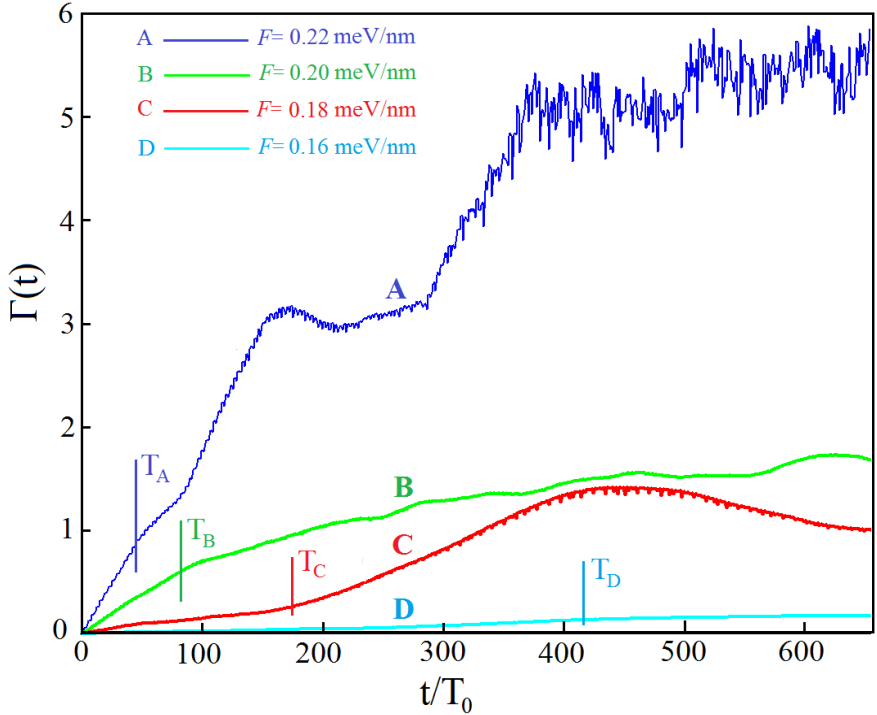}\hspace{2pc}%
\begin{minipage}[b]{50mm}\caption{\label{figint} Integrated ionization rates $\Gamma(t)$ from (\ref{gamma}) obtained for numerical results from \cite{KLS2020} for different driving field strength $F$ labeled by $A,B,C,D$. The analytical ionization times are labeled as $T_A,T_B,T_C,T_D$.}
\end{minipage}
\end{figure}

The integrated rate $\Gamma(t)$ obtained with numerical data from \cite{KLS2020} is shown in Fig.\ref{figint} for the parameters described in previous Section and is shown for four driving field amplitudes $F=0.16,..,0.22$ meV/nm labeled by curves A,...,D, respectively. The labels $T_A,...,T_D$ are the analytically predicted ionization times that will be discussed below. For constant ionization rate the integrated rate $\Gamma(t)$ would be a linear function of time, but the curves in Fig.\ref{figint} clearly show that the actual ionization develops with a varying rate $w(t)$ which slows with time. 
The numerical ionization time can be extracted from the points where $\Gamma(t) \approx 1$, and from  Fig.\ref{figint} we can see that numerical results predict ionization times that are typically longer than analytical ones. The best fit is obtained for the strongest field $F=0.22$ meV/nm while for the lowest field $F=0.16$ meV/nm the point $\Gamma(t) \approx 1$ was never reached on the interval of observation, indicating that for low field the electron may stand in a confined state longer than analytically predicted.

The analytical description of ionization rate $w_a$ and associated ionization time $T=1/w_a$ is performed using the results for finite-width square quantum well from \cite{Demi1996} adjusted for finite wire length:

\begin{equation}
w_a=\frac{A e^{\kappa d}}{1+\frac{1}{2}\kappa d} 
\left(1-\frac{|E_1|}{U_0} \right) w_{\delta},
\label{wa}
\end{equation}

where $\kappa=1/l$ and $l\approx 25$ nm is our numerically obtained decay length for the ground state wavefunction in \cite{KLS2020}, $A$ is a finite-wire factor described below, and $w_{\delta}$ is the ionization probability per unit time for the 1D well with delta-function confining potential. In the limit when the Keldysh adiabaticity parameter \cite{DK1985,DK2000,Keldysh1964} 
$\gamma=\hbar \omega \kappa/F$ is lower than $1$ (in our problem 
$\gamma \approx 0.2...0.3$) one can use the tunneling approximation giving for $w_{\delta}$ \cite{Perelomov1966,Demi1996}

\begin{equation}
w_{\delta} \approx 2\frac{|E_1|}{\hbar} \left(\frac{3F}{\pi F_0} \right)^{1/2}
\exp \left[-\frac{2 F_0}{3F} \right],
\label{wdelta}
\end{equation}

where $F_0=2\kappa |E_1|$ is a typical barrier electric field equal to $1.82$ meV/nm in our model.

The finite wire factor $A$ in (\ref{wa}) reflects the fact that for a finite nanowire the long-time outcoming flux $j_{+}$ defining the ionization probability $w_{a}=2 j_{+}$ \cite{Demi1996} can be modified due to numerous reflections from the wire leads. Assuming good reflection with flux coefficient $R \approx 1$ and neglecting the interference effects due to arbitrary point $x$ where the flux is calculated one may obtain by various summation rules (geometric progression, time averaging, etc.) that $A \to 1/2$ in the limit of infinite reflection sequence. With such approximation the equations (\ref{wa}), (\ref{wdelta}) give the ionization rates and associated ionization times $T=1/w_a$ that are marked by labels $T_A,..,T_D$ in Fig.\ref{figint}.

\begin{figure}[h]
\includegraphics[width=115mm]{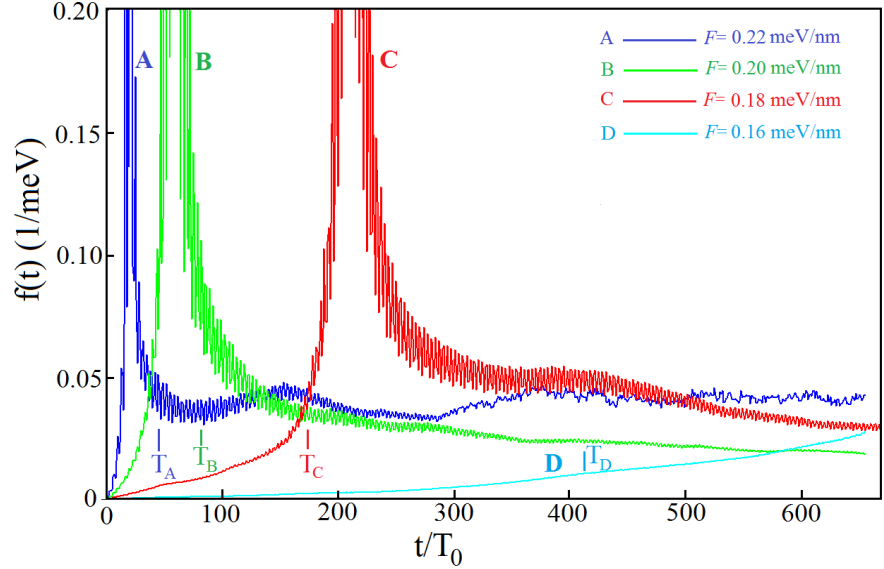}\hspace{2pc}%
\begin{minipage}[b]{50mm}\caption{\label{figfit} Plots for $f(t)$ from (\ref{ft}) connecting the integrated ionization rate (\ref{gamma}) and mean energy $\langle E(t) \rangle$. Peaks of $f(t)$ satisfactory correspond to analytical ionization times $T_A,T_B,T_C$.}
\end{minipage}
\end{figure}

By analyzing the numerical results for $\Gamma(t)$ we have noticed that the plots in Fig.\ref{figint} resemble the time dependencies of mean energy $\langle E(t) \rangle$ that were obtained in \cite{KLS2020} for the same parameters. We have found in \cite{KLS2020} that the crossing of threshold $\langle E(t) \rangle=0$ between the localized and continuum states may serve as an indication of effective tunneling time. In Fig.\ref{figfit} we plot the function $f(t)$ connecting the integrated ionization rate $\Gamma(t)$ from (\ref{gamma}) and mean energy  $\langle E(t) \rangle$:

\begin{equation}
f(t)=\frac{\Gamma(t)}{\langle E(t) \rangle}.
\label{ft}
\end{equation}

One can see from Fig.\ref{figfit} that the peaks of $f(t)$ reflecting the threshold crossing $\langle E(t) \rangle \approx 0$ are in satisfactory agreement with analytical ionization times $T_A,T_B,T_C$ obtained from  (\ref{wa}), (\ref{wdelta}). The difference between the peak position and analytical estimate is about $15 \%$ for $F=0.20$ meV/nm, $20 \%$ for $F=0.18$ meV/nm, and $35 \%$ for $F=0.22$ meV/nm, which is considered as a good degree of correspondence for ionization problems \cite{Bauer1999}. The only exclusion in Fig.\ref{figfit} is for the lowest field $F=0.16$ meV/nm where no peak is observed since here the mean energy never crosses the threshold $\langle E(t) \rangle=0$ on the observation interval, i.e. the electron remains confined in the well. For long times $t\gg T_{A,B,C}$ the plots for $f(t)$ approach almost the same plateau indicating the proportionality of the final integrated rate (\ref{gamma}) to  $\langle E(t) \rangle$, i.e. in classical approximation to $F^2$. Similar fitting has been observed for atomic ionization problem \cite{Bauer1999}. So, for low fields the numerical approach differs from analytical predictions since here the role of confining potential is more pronounced while for higher fields the two approaches demonstrate satisfactory agreement.

\section{Conclusions}
We have made connections between the numerical results for the driven evolution of the electron in a gated nanowire with spin-orbit coupling in the EDSR regime \cite{KLS2020} where the tunneling to the continuum takes place and the analytical approximations for the quantum well ionization rate. It is found that the numerical ionization rate is time-dependent, and the analytical ionization times are in satisfactory correspondence with the numerical threshold times where the mean energy crosses the border between the localized and continuum states. We believe that our findings will help in clarification of the results for various numerical modeling of the electron driven dynamics in nanostructures.

\section*{Acknowledgements}
The author is grateful to E.Ya. Sherman for very valuable discussions.
The work is supported by the Ministry of Higher Education and Science of Russian Federation under the Project part of State assignment 2020.

\end{document}